\RequirePackage{lineno}
\documentclass[
reprint,
twocolumn,
secnumarabic%
,amssymb,
nobibnotes, nofootinbib, aps, prl, showpacs,showkeys]{revtex4}
\usepackage{docs}%
\usepackage{bm}%
\usepackage{graphicx}
\usepackage{hyperref}
\usepackage{url}

\expandafter \ifx \csname package@font\endcsname\relax\else
\expandafter \expandafter \expandafter
\usepackage
\expandafter \expandafter
\expandafter{\csname package@font\endcsname}%
\fi



\usepackage[normalem]{ulem}

\usepackage{epstopdf}

\begin{document}

\setlength\columnsep{25pt}

\author{Tomasz Golan}
\author{Krzysztof M. Graczyk}
\author{Cezary Juszczak}
\author{Jan T. Sobczyk}
\affiliation{{ }\\
Institute of Theoretical Physics, Wroc\l aw University\\
Plac Maxa Borna 9, 50-204 Wroc\l aw, Poland}

\title{Extraction of Axial Mass and Strangeness Values from the MiniBooNE 
Neutral Current Elastic Cross Section Measurement}

\begin{abstract}
Results of the analysis of the MiniBooNE experiment data for the neutral
current elastic neutrino scattering off the $CH_{2}$ target with the NuWro Monte
Carlo generator are presented.  Inclusion in the NuWro the two body current 
contribution leads
to the axial mass value $M_A =  1.10^{+0.13}_{-0.15}$~GeV consistent with the 
older evaluations based on the
neutrino-deuteron scattering data. The strange quark contribution
to the nucleon spin is estimated with the value $g_A^s = -0.4^{+0.5}_{-0.3}$.
\end{abstract}

\maketitle

\section{Introduction}
\label{sec_introduction}

There has been a lot of interest in neutrino interactions in the few GeV energy 
region, coming from the
oscillation experiments and the demand to better constrain the systematic 
errors.
For the neutrino energies around $1$~GeV the most abundant reaction is
charged current quasi-elastic (CCQE) scattering: $\nu_\mu + n
\rightarrow \mu^- + p$ and it is also the most important process in the investigation of the
oscillation phenomenon, e.g. in the T2K experiment \cite{t2k}. 

Due to the standard conserved vector current (CVC) and partially conserved 
axial 
current (PCAC) hypotheses, using the electron scattering data and assuming 
a dipole form of the axial form factor, the weak transition matrix element contains only one unknown parameter, the axial mass $M_A$.
Recent neutrino CCQE cross section measurements, notably the high statistics muon double
differential cross section results from the MiniBooNE (MB) experiment
\cite{MB-CCQE}, suggest $M_A$ values significantly larger than
estimations from the older deuteron target neutrino measurements
and from the pion electro-production data \cite{Bernard:2001rs}. It is becoming 
clear that
in order to understand correctly the MB data, a two body current contribution 
to 
the cross section must be considered \cite{MEChM, CCQE,
NR-SVV-1, BBC}. In the experimental event 
identification this contribution can easily be confused with the 
genuine CCQE events. The older and recent CCQE $M_A$ estimates can be 
consistent, because in the case of the neutrino-deuteron scattering the two body
current contribution is small \cite{shen}. 

Theoretical models of the two body current contribution (called in this paper also n-particles n-holes ($np-nh$))
give quite different
estimates of the size of the effect both for neutrino and antineutrino 
scattering. Recently the MB Collaboration
published the first large statistics antineutrino CCQE-like (a sum of CCQE and 
$np-nh$ contributions)
cross section results \cite{MBanti}. The data has been already analysed by the 
theoretical groups \cite{Nieves:2013fr, Martini_anti}.
In Ref. \cite{JoGra} a ratio of the CCQE-like cross
sections for neutrinos and
antineutrinos was discussed as a function of neutrino energy. If the errors become smaller, this kind of data can allow to discriminate the
models. It is important to look also for alternative ways to evaluate $M_A$ and/or investigate the size
of the two body current contribution. The MINERVA Collaboration analysed the 
energy deposit near CCQE-like interaction vertex looking 
for an evidence for multinucleon knock-out events \cite{MINERVA}. 

Another option is
to look at the neutral current elastic (NCEL) reaction: $\nu_l +
N\rightarrow \nu_l + N$, where $N$ denotes proton or neutron. As it will be 
explained in detail in Sect. \ref{ncel} the basic theoretical framework to
investigate NCEL scattering is similar to the one used for CCQE. The neutral 
current NC
hadronic current is expressed in terms of the vector and axial
form factors, that are linear combinations of the form factors
present in the CCQE reaction with an addition of new terms sensitive to the 
strange quark content
of nucleons. Thus, the NCEL scattering 
data allows for extraction of both $M_A$ and the 
strange quark contribution to the form factors. In fact, most of the interest in 
the NCEL
reaction comes from its potential to measure the strangeness of the nucleon.


The axial strange form factor can be determined
from the NCEL neutrino-$CH_2$ scattering data because it enters proton and 
neutron matrix elements with
opposite signs. Typically, in experiments, one tries to extract
$g^s_A$, a value of the axial strange form factor at $Q^2=0$. It corresponds to 
the fraction of strange quarks
and antiquarks that contribute to the total proton spin, commonly denoted as $\Delta s$. 
The first estimation based on $\nu N$ data (BNL E734
experiment) was done in Ref. \cite{Ahrens:1986xe} with the result $g_A^s =
-0.15\pm0.09$. For the later discussions see Refs. \cite{GLW,
ABBCGMMdGU}.


About three years ago the MB Collaboration measured the flux averaged NCEL 
differential cross section on the $CH_2$
target \cite{MiniBooNE_AguilarArevalo:2010cx}.
Two observables were considered. The first one is the distribution of events 
in the total reconstructed kinetic energy of the final state nucleons. The 
measurement is based on
the MB detector ability to analyse the scintillation light in the absence
of Cherenkov light from the final state muon. In the second observable 
a proton enriched sample of
events is analysed with the kinetic energies above
the Cherenkov radiation threshold. The observable is defined as the ratio of 
the cross section for the proton enriched sample to the total NCEL-like (events 
with no pions in the final state) cross section.
The MB Collaboration made two separate parameter extractions.  Assuming $g_A^s=0$ the value of $M_A^{eff}=1.39\pm 0.11$~GeV was obtained from the first observable.
Taking the value $M_A^{eff}=1.35$~GeV from the CCQE analysis \cite{MB-CCQE}, MB found 
$g_A^s =0.08\pm 0.26$ from the proton enriched sample of events. The description of both data samples in terms of 
$M_A^{eff}\sim 1.35$~GeV and $g_A^s \sim 0$ is consistent. The MB Collaboration did not attempt to make a simultaneous fit to both
theoretical parameters (such fits were only discussed in Ref. 
\cite{Per-thesis}).

In the MB analysis the values of $M_A$ used in modelling scattering off carbon 
and off free protons, were different.
For protons the $M_A$ was fixed to be $1.13$~GeV. For carbon the axial mass values was treated as a free parameter.
A large {\it effective} $M_A$ value is expected to account for the $np-nh$ 
contribution present in
neutrino-carbon scattering but absent in the neutrino-proton
scattering. MB Collaboration used NUANCE neutrino event generator with nuclear 
effects described by Fermi gas model and final state interactions (FSI) 
\cite{NUANCE}. NUANCE does 
not include the
$np-nh$ contribution.

The data from Ref. \cite{MiniBooNE_AguilarArevalo:2010cx} has been already 
discussed 
in 
some phenomenological papers.
Butkevich and Perevalov \cite{BuPe} investigated an impact of more realistic nuclear model
in the MB data analysis. A relativistic distorted wave
impulse approximation (RDWIA) \cite{BuKu} model leads to the values $M_A=1.28\pm 
0.05$~GeV and $g_s^A=-0.11\pm 0.36$, consistent with those reported in Ref.
\cite{MiniBooNE_AguilarArevalo:2010cx}. The spectral function
formalism was used by Ankowski in Ref. \cite{Ank-NC}, with the conclusion that 
the shape of the MB measured
distribution of events in $Q^2$ is reproduced with $M_A^{eff}=1.23$~GeV.
However, there is a $20\%$ discrepancy in the overall
normalisation with the MB results (the measured cross section is larger). 
Meucci,
Gusti and Pacati analysed the predictions of four nuclear models
\cite{MGP}. It turned out that relativistic Green function (RGF)
model is able to reproduce the MB NCEL data with the value of $M_A$ close to 
those
obtained in old deuteron-target experiments. 

The purpose of this paper is to perform the first analysis of the MB NCEL data 
using a model 
that includes the  $np-nh$ contribution. The model is provided by the NuWro 
Monte Carlo event 
generator \cite{GJS} developed over last 9 years at the Wroc\l aw University.
NuWro describes the neutral current $np-nh$ contribution with the effective
transverse enhancement (TE) 
model \cite{BBC}.
The TE model provides the contribution to the neutrino inclusive cross section and  
predictions for the final state nucleons are obtained with a procedure 
described in Ref. \cite{So-MEC}. 
The advantage of the TE model is that it can be safely used for neutrinos of 
energies 
larger than $1.5$~GeV.
In this paper only nucleons resulting from the $np-nh$ events will be analysed 
and differences between microscopic (see Refs. \cite{NR-SVV-1, MEChM}) and 
effective $np-nh$ 
models are made smaller by Fermi motion and final state interactions effects. 
The recent MINERVA CCQE-like data analysis demonstrates that with 
$M_A=0.99$~GeV and 
the TE model both neutrino and antineutrino $Q^2$ distributions are well 
reproduced \cite{MINERVA}.

Another important difference with respect to the previous studies of the MB 
NCEL data
is in the treatment of nuclear effects. We will compare NuWro predictions to 
the quantities (visible energy) that are directly observable and
we will not rely on the NUANCE FSI model. 

Our main result is a simultaneous fit to the $M_A$ and $g_A^s$ done using the 
MB data for the distribution of events in the total reconstructed kinetic 
energy of the final state nucleons, with the outcome: $M_A =  
1.10^{+0.13}_{-0.15}$~GeV and $g_A^s = -0.4^{+0.5}_{-0.3}$. We will argue that 
the second MB observable is very sensitive to details
of FSI and the $np-nh$ kinematics model and it is very difficult to use it as a 
reliable source of 
an information about theoretical model parameters. On the 
contrary, the first observable is quite robust to such details and can be 
successfully used to 
extract values of the interesting quantities.

Our paper is organised as follows: in Sec. \ref{theory} a general description 
of the NCEL reaction is given;
in Sec. \ref{nuwromc} the main features of the NuWro generator are summarised;
Sec. \ref{dataanal} contains a details of the data analysis and the MB energy unfolding procedure;
in Sec \ref{results} we present our main results, and conclusions can be found in Sec. \ref{conclusions}.


\section{Elastic Neutral Current Reaction Formalism}
\label{theory}
\label{ncel}

We consider neutral current neutrino-nucleon scattering:
\begin{equation}
\nu(k) + N(p) \to l'(k') + N'(p'),
\end{equation}
where $N$, $N'$, $l'$ denote the initial and final state nucleons and the 
outgoing lepton with the four momenta: $p$, $p'$ and $k'$, respectively. 
The four momentum transfer is given by $q^\mu \equiv k^\mu - {k'}^\mu = (\omega, \mathbf{q})$,
$Q^2 \equiv -q^2$.

In the Born approximation the scattering matrix element factorises into the 
product of leptonic and hadronic contributions:
\begin{equation}
i\mathcal{M}^{(1)}_{nc} \approx - i 
\frac{G_F}{\sqrt{2}} j_\mu h^\mu_{nc}, \quad Q^2 \ll 
M_Z^2,
\end{equation}
where $\theta_C$ is Cabibbo angle, $G_F$ is the Fermi constant, while $j_\mu$ 
and
$h^\mu_{cc,nc}$ are the expectation values of the leptonic and hadronic 
currents. The leptonic part is:
\begin{eqnarray}
j_\mu &=& \bar{u}(k') \gamma^\mu(1-\gamma_5) u(k),
\end{eqnarray}
while getting the hadronic contribution requires an extra phenomenological input in the general formula
\begin{equation}
h^\mu(q) = \overline{u}(p') \Gamma^\mu(q)  u(p),
\end{equation}
with the effective hadronic vertex $\Gamma^\mu$.

In order to construct $\Gamma^\mu$ for the NCEL 
scattering, one has to follow the pattern given by the
Standard Model. The CVC theory, the PCAC hypothesis and the
$SU(2)$ isospin symmetry relate the neutral current form factors to those present in the electromagnetic 
and charged current hadronic matrix elements \cite{Alberico:2001sd}. The NC 
hadronic vertex reads:
\begin{widetext}
\begin{equation}
\label{NCvertex} \Gamma^\mu_{NC,p (n)} = \gamma^\mu F_1^{NC,p(n)}
+ \frac{i \sigma^{\mu\nu}q_\nu}{2M} \gamma^\mu F_2^{NC,p(n)} -
\gamma^\mu \gamma_5 G_A^{NC,p(n)},
\end{equation}
where indices $p$ and $n$ refer to proton and neutron. The NC form factors can be expressed as:
\begin{eqnarray}
\label{eq: ff}
F_{1,2}^{NC,p(n)} (Q^2) &=& \pm \frac{1}{2} \left\{F_{1,2}^p(Q^2)
- F_{1,2}^n(Q^2)\right\} - 2 \sin^2\theta_W F_{1,2}^{p(n)}(Q^2)  -
\frac{1}{2} F_{1,2}^s(Q^2),
\\ \label{eq: ff2}
G_A^{NC,p(n)} (Q^2)&=& \pm \frac{1}{2}G_A(Q^2) -
\frac{1}{2}G_A^s(Q^2),
\end{eqnarray}
\end{widetext}
$+/-$ signs refer to proton/neutron, $\theta_W$ is the Weinberg angle, $\sin^2\theta_W = 0.231$. $F_{1,2}^{p(n)}$ 
are the proton (neutron) electromagnetic form factors, $G_A$ is the axial 
nucleon form factor:
\begin{equation}
\label{axial_form_factor} G_A(Q^2) =
\frac{g_A}{\left(1 + \frac{Q^2}{M_A^2}\right)^2}, \qquad g_A=1.267.
\end{equation}
$F_{1,2}^s$ and $G_A^s$ are the vector and the axial strange form factors.

The electromagnetic form factors $F_1$, $F_2$ are obtained from the analysis of
the elastic $eN$ scattering data (for a review see Ref. \cite{Perdrisat:2006hj}) 
and can be expressed in terms of the electric and magnetic form factors:
\begin{eqnarray}
F_1^{p(n)} (Q^2) &=& \nonumber \\
& & \!\!\!\!\!\!\!\!\!\frac{4M^2}{Q^2 +4M^2} \left[
G_E^{p(n)}(Q^2) + \frac{Q^2}{4M^2} G_M^{p(n)}(Q^2)\right],
\nonumber
\\
\\
F_2^{p(n)} (Q^2) &=& \frac{4M^2}{Q^2 +4M^2} \left[
G_M^{p(n)}(Q^2) -  G_E^{p(n)}(Q^2)\right] ,
\end{eqnarray}
where $M = \frac{1}{2}(M_p + M_n)$ is the average nucleon mass.

In the Breit frame the electric ($G_E^{p(n)}$) and the
magnetic ($G_M^{p(n)}$) nucleon form factors are related to the nucleon
electric charge and current distributions. For instance, for the electric proton form factor $G_E^{p}(Q^2) \sim
G_E^{p}(0) + \frac{\langle r_p^2 \rangle}{6} Q^2 + O(Q^4)$, where $\langle r^2_p \rangle = -\left. 6\frac{d
G_{E}^p}{dQ^2} \right|_{Q^2=0}$ is the mean-square radius of the charge distribution.

Similarly, one can introduce the electric ($G_E^s$) and magnetic
($G_M^s$) isoscalar strange form factors. In the
first approximation one can assume that ``the effective sizes'' of
the proton and neutron strange sea are the same, therefore
we use the same strange form factors for both nucleons. The latest global 
analysis of the elastic parity violating (PV) $ep$ and BNL E734 
neutrino scattering data 
indicate that the 
values of vector strange form
factors are consistent with zero \cite{Pate:2010kz}, and in our analysis we set $G_{E,M}^{s}=0$.

In order to estimate the axial strange form factor we assume
that the radius of the strange sea is comparable with the axial
``charge'' radius of the proton so that 
\begin{equation}
\label{axial_strange_form_factor} G_A^s(Q^2) =
\frac{g_A^s}{\left(1 + \frac{Q^2}{M_A^2}\right)^2},
\end{equation}
with the free parameter $g_A^s$ that has to be extracted from the data.

In our numerical analysis we use the BBBA05 vector form factors \cite{bbba05}.
We also investigated a possible impact on the results from the form factors 
corrected 
by the two-photon exchange effect
\cite{Graczyk:2011kh}.

\section{NuWro Monte Carlo event generator}
\label{nuwromc}
\subsection{Generalities}

NuWro is a Monte Carlo event generator developed at the Wroc{\l}aw University 
\cite{GJS}.
It simulates neutrino-nucleon and neutrino-nucleus interactions including: 
(quasi-) elastic scattering, pion production through $\Delta(1232)$ resonance 
(with a contribution from the non-resonant background), 
more inelastic processes and coherent pion production. 
NuWro covers neutrino energy range from $\sim100$~MeV to TeV.
There are three basic nucleus models implemented in NuWro: global/local relativistic
Fermi gas (FG) model, spectral function and the effective momentum and density 
dependent potential. In the analysis described in this paper we use the 
local Fermi gas model.
In the impulse approximation picture,
all the hadrons arising at a primary vertex are propagated through the nuclear 
matter using the NuWro cascade model. NuWro is the open-source project and the 
code is freely available \cite{repo}.

\subsection{Two Body Current Contribution}
\label{sec: mec}

\begin{figure}
\centering{
\includegraphics[width=\columnwidth]{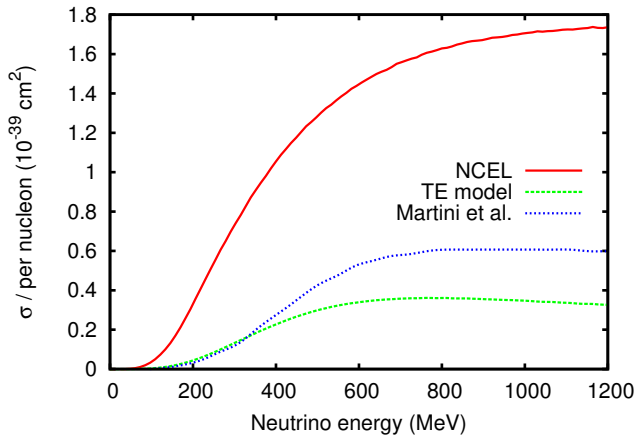}
\caption{[Color online] Total cross section per nucleon as a function of a 
neutrino energy for NC scattering off carbon: NuWro model with $M_A = 
1.03$~GeV and Martini et al. \cite{MEChM}.}
\label{fig: mec_nc}}
\end{figure}

There are three charged current (CC) two-body current models implemented in 
NuWro: the IFIC model 
\cite{NR-SVV-1}, the Martini et al. model \cite{MEChM} and the TE model 
\cite{BBC}.
In all of them double differential cross section contribution for the final state muons is the external input
and the hadronic part is modelled using the scheme (proposed in \cite{So-MEC}):

\begin{enumerate}
 \item set randomly the four-momenta ($p_1$ and $p_2$) of the initial nucleons from the Fermi sphere with a radius determined by the local nuclear density;
 \item calculate the four-momentum of the hadronic system $$W = p_1 + p_2 + q,$$ 
where $q$ is the four-momentum transferred to the hadronic system;
 \item repeat steps 1.\ and 2.\ until the invariant hadron mass is larger than 
mass of two nucleons ($W^2 > (2M)^2$);
 \item make the Lorentz boost to the hadronic center of mass system;
 \item select isotropically momenta of two final state nucleons;
 \item boost back to the laboratory frame;
\item apply final state interactions.
\end{enumerate}

Results from three models depend on the distribution of events in the energy 
transfer. The differences are smeared out by the Fermi motion and the FSI 
effects. 

In our analysis of the NCEL data we use the TE model \cite{BBC}, because to 
date this is the only NC $np-nh$ model available in NuWro. 
In the TE model the $np-nh$ contribution to the NC
scattering cross section on carbon is introduced by the modification of the
vector magnetic form factors:
\begin{equation}
 G_M^{p,n} \rightarrow \tilde{G}_M^{p,n} = \sqrt{1 + AQ^2\exp{\left(-\frac{Q^2}{B}\right)}} G_M^{p,n}(Q^2),
\end{equation}
where $A = 6~\mathrm{GeV}^{-2}$ and $B = 0.34~\mathrm{GeV}^2$. Using $\tilde{G}_M^{p,n}$
in the NCEL differential cross section formula one obtains
a cross sections for a sum of the NCEL and $np-nh$ reactions.
The $np-nh$ cross section is obtained by subtracting the NCEL part described with the standard magnetic form factors:
\begin{eqnarray}
 \frac{d^2\sigma^{\mathrm{MEC}}}{dqd\omega} & \equiv &
 \frac{1}{2}\left\{\left(\frac{d^2\sigma^{\mathrm{NCE}}}{dqd\omega}(\tilde{G}_M^{p}) - \frac{d^2\sigma^{\mathrm{NCE}}}{dqd\omega}(G_M^{p})\right)\right.
 \nonumber \\ & + & \left.\left(\frac{d^2\sigma^{\mathrm{NCE}}}{dqd\omega}(\tilde{G}_M^{n}) - \frac{d^2\sigma^{\mathrm{NCE}}}{dqd\omega}(G_M^{n})\right)\right\}.
\end{eqnarray}

The value of the axial mass in the TE model is set to be
$M_A^{np-nh} = 1014$ MeV, as assumed in Ref. \cite{BBC}. The total cross 
section for the 
NC $np-nh$ scattering is shown in Fig. \ref{fig: mec_nc}. 
At the typical MB flux neutrino energies, $E_\nu \sim 700$~MeV, the $np-nh$ 
contribution in the TE model amounts to about $19\%$ of the NCEL cross section. 
For comparison we show also the predictions from the Martini et al. model taken 
from Ref. \cite{MEChM}. 
The Martini et al. model predicts much larger $np-nh$ cross section. Also for 
the CC $np-nh$ reaction the predictions from this model are 
larger than those from the IFIC and TE models.

In the CC $np-nh$ reactions there can be either neutron-neutron ($n$-$n$) or proton-neutron ($n$-$p$) pair in the initial state.
The probability of the mixed isospin pair is defined in NuWro by a parameter 
$p_{CC}$, with the default value $p_{CC} = 0.6$.
For the NC $np-nh$ interactions every isospin initial state pair is possible ($n$-$p$, $n$-$n$, $p$-$p$).  To keep the same proportion between 
$n$-$n$ and $p$-$n$ pairs  and assuming the same probability to have $n$-$n$ and $p$-$p$ pairs, we introduce the parameter $p_{NC} = (2/p_{CC} - 1)^{-1}$
giving the likelihood of a $n-p$ pair to be selected in a NC two body current reaction. 

\subsection{NuWro Cascade Model}
\label{sec: nucmodel}

Hadrons resulting in the primary vertex propagate through the nuclear matter 
with 
the NuWro cascade model:

\begin{enumerate}

 \item nucleons are assumed to be in the potential well of depth $V = V_0 + E_F$, where $E_F$ is the Fermi energy and $V_0 = 7$~MeV;
 \item formation zone (FZ) can be applied (FZ is set to be zero for the $np-nh$ events) and the following steps are repeated:
\begin{enumerate}
 \item a nucleon free path ($\lambda$) is drawn from the exponential 
distribution taking into account the nucleon-nucleon cross section 
 and the local nuclear density;
 \item if $\lambda \leq\lambda_{max} = 0.2$~fm the nucleon is propagated 
by 
$\lambda$, the interaction kinematics is generated 
 and a check for Pauli blocking is done to decide if the interaction happened;
 \item if $\lambda >\lambda_{max}$ the nucleon is propagated by 
$\lambda_{max}$.
\end{enumerate}
\end{enumerate}

In the cascade the Fermi motion of target nucleons is taken into account, and every new 
nucleon, which participate in the cascade, brings in an extra kinetic energy. 
When a nucleon leaves the nucleus, its kinetic energy is reduced by $V$ 
setting it on-shell. If the kinetic energy is smaller than $V$, the nucleon is assumed to be stuck inside the nucleus.

For more detailed description of the NuWro cascade model see Ref. \cite{GJS}.

\section{Data Analysis}
\label{dataanal}

We are going to discuss two data samples provided by the MB Collaboration.
The first one (the NCEL sample) contains the distribution of the total 
reconstructed kinetic energy of all nucleons in the final state, normalised to 
the number of events seen in 
the detector.

The second data sample (the NCEL high energy sample) is provided in a form 
of the ratio:
\begin{equation}
  \eta = \frac{\tilde{X}(\nu p \rightarrow \nu p)}{X(\nu N \rightarrow \nu N)},
  \label{eq: ratio}
\end{equation}
where $\tilde{X}$ denotes a contribution from a special class
of events, called single proton or proton enriched. Those are the events with
visible Cherenkov light and proton angle
$\theta < 60^0$. In MC simulations the largest contribution to those events 
comes
from the NCEL scattering on protons, which then do not undergo
reinteractions. In the case of multiple proton events, the energy
of individual proton is in general too low to produce the Cherenkov
light. Even if a high energy proton appears in the multiple proton
event, it has typically larger scattering angle than protons unaffected by
FSI. The denominator ($X$) denotes the contribution from all the NCEL-like 
interactions.

Both data samples are presented as a function of reconstructed
energy\footnote{We keep the original notation from \cite{Per-thesis}.}
($\nu$), measured in the detector. To compare
those data with the theoretical predictions given in terms of the true
kinetic energy ($\mu$), one needs the unfolding procedure, allowing a passage 
from $\mu$ to $\nu$.

In the next subsections we describe the original MB unfolding procedure. 
We had to propose a treatment of $np-nh$ events, not considered in the MB 
analysis. 

\subsection{MiniBooNE Procedure}
\label{anal1}

For all but $np-nh$ events we follow closely the approach proposed by Perevalov in Ref. \cite{Per-thesis}.
A similar unfolding procedure is used in the both data samples. Five types of 
events giving
contribution to the final distributions are considered:

\begin{enumerate}

 \item[](a) NCEL on hydrogen;
 \item[](b) NCEL on a proton from carbon unaffected by FSI effects;
 \item[](c) NCEL on a proton from carbon with FSI effects;
 \item[](d) NCEL on a neutron from carbon;
 \item[](e) irreducible background (pions produced in a primary vertex and absorbed due to FSI effects).

\end{enumerate}

The probabilities of scenarios (b-e) depend on the details of the cascade 
models implemented in
NUANCE or in NuWro.

For each type of the signal events, $k=1,2,...,5$,  there is a
response matrix ($R^{(k)}$)  provided in Ref. \cite{MB-page}, which 
simulates the energy
smearing, the detector efficiency and defines a relation between true and reconstructed energy
distributions:

\begin{equation}
\nu_j^{(k)} = \sum_{i} R_{ij}^{(k)} \mu_i^{(k)}.
\label{eq: t2r}
\end{equation}

$R^{(k)}$ are either $51 \times 51$ or $30\times 30$ matrices for the two data 
samples, respectively. The columns of
matrices label the true kinetic energy bins and rows label the
reconstructed energy. There are $50$ bins starting from
$0$~MeV up to $900$~MeV plus an extra overflow bin for the NCEL
sample in the true kinetic energy. For NCEL high energy sample there
are $28$ bins, starting from $300$~MeV up to $900$~MeV plus
underflow and overflow bins.

To obtain the reconstructed kinetic energy distribution and compare 
with the data from Ref. \cite{MiniBooNE_AguilarArevalo:2010cx} one 
goes through the following steps:

\begin{enumerate}
 \item use a theoretical model and calculate the flux-averaged distributions for five different types of signal events using the same bins 
 as in the response  matrices;
 \item use the proper response matrices to translate each histogram to the reconstructed kinetic energy distribution;
 \item sum all the histograms and add the background events ($\nu^{BKG}$ 
contains dirt, beam-unrelated, and other backgrounds 
 provided by the MB Collaboration in Ref. \cite{MB-page}) to get the total 
reconstructed energy spectrum:

\begin{eqnarray}
 \nu_j^{MC} & = & \sum_i R_{ij}^{(1)} \mu_i^{(1)} + \sum_i R_{ij}^{(2)} \mu_i^{(2)} \nonumber \\
        & + & \sum_i R_{ij}^{(3)} \mu_i^{(3)} + \sum_i R_{ij}^{(4)} \mu_i^{(4)} \\ \nonumber
        & + & \sum_i R_{ij}^{(5)} \mu_i^{(5)} + \nu_j^{BKG}
\end{eqnarray}

 \item use the provided error matrices ($M_{ij}$) to calculate $\chi^2$:

\begin{equation}
 \chi^2 = \sum_i \sum_j \left(\nu_i^{DATA} - \nu_i^{MC}\right) M_{ij}^{-1} \left(\nu_j^{DATA} - \nu_j^{MC}\right)
 \label{eq: chi2}
\end{equation}

Unlike in the CCQE MB data published in Ref. \cite{MB-CCQE}, the flux 
normalisation 
error is already included in the error matrix $M_{ij}$.

\end{enumerate}

\subsection{Our Procedure}
\label{anal2}

An alternative way to convert the true kinetic energy to the reconstructed one is to translate it on the event by event basis. 
For each value of the true kinetic energy the corresponding column in the response matrix
gives a probability distribution with the information how the given true energy 
value is smeared out in the detector,
normalised to the efficiency. To obtain the reconstructed kinetic energy 
distribution one proceeds as follows:

\begin{enumerate}

 \item for each event calculate the total true kinetic energy of all nucleons 
in the final state ($\mu$) and get a bin number $j$;
 \item find the type of signal ($k$), see Sect. \ref{anal1};
 \item choose $j$-th column of the $R^{(k)}$ response matrix as the probability distribution;
 \item use the MC method to decide if the event is accepted (according to 
the efficiency)
and what energy is visible in the detector.

\end{enumerate}

\subsubsection{The Unfolding Procedure for Two Body Current Events}

In the $np-nh$ events there are typically two nucleons after a primary interaction 
and both of them propagate independently through nucleus. 
In the MB analysis there are no $np-nh$ events included and no response matrices were prepared for them. To take two body current events into account, 
we must express them in terms of the five signals defined by the MB.

A naive interpretation may suggest a treatment of each nucleon from the $np-nh$ events separately. However, it would be incorrect. 
The signal is recorded by photomultiplier tubes (PMT), which absorb the light emitted in the scintillator (and also the Cherenkov radiation). 
The event is accepted if there is a sufficient number of PMT hits. Any of two individual nucleons may have too low energy to generate enough PMT hits, but together they can make it \cite{Perevalov_discussion}.

In our analysis we treated both nucleons from a two body current event together and summed up the kinetic energies of all the nucleons in the final state 
as if they were coming from only one nucleon. In the detector $np-nh$ events 
are seen as multiple protons events, so as signals (c) - NCEL on proton from 
carbon affected by FSI effects and (d) - NCEL on neutron from carbon.

One expects events without a proton 
in the final state to be more smeared out in the detector, and we apply response 
matrix for the signal (d) if there were two neutrons in the primary vertex or 
for the signal (c) in other cases 
\cite{Perevalov_discussion}.

In $np-nh$ events the energy transferred to the hadronic system is shared by two 
nucleons,
and the probability that there will be a proton with energy large enough to 
produce the Cherenkov light is low. Thus, the $np-nh$ events make the ratio 
$\eta$ (Eq. \ref{eq: ratio}) smaller.


\section{Results}
\label{results}

\begin{figure}
\centering{
\includegraphics[width=\columnwidth]{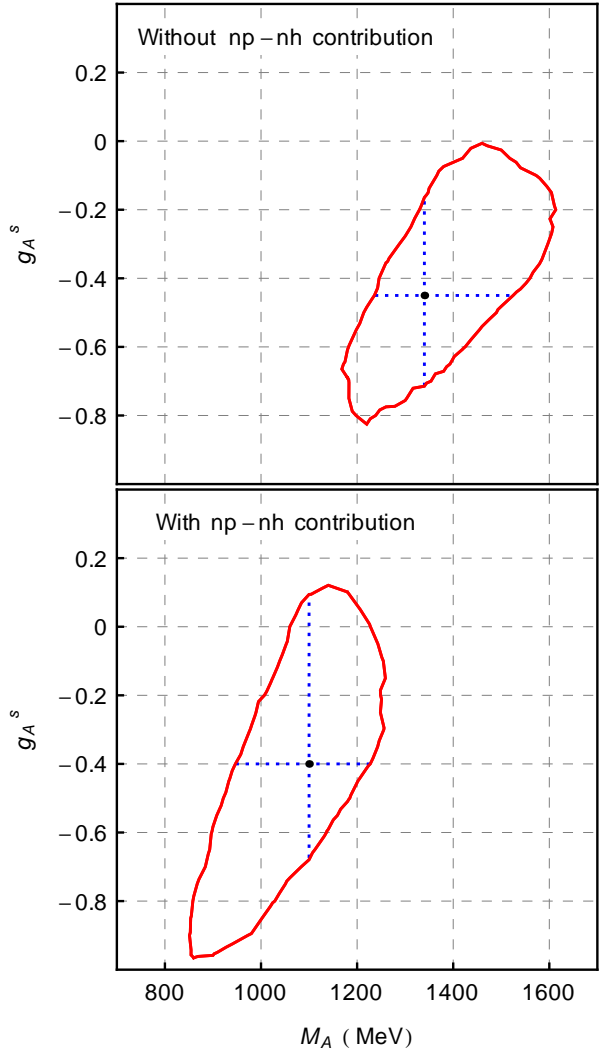}
\caption{[Colour online] $1\sigma$ error contour for
$(M_A,g_A^s)$ parameters obtained from $\chi^2$ (Eq. \ref{eq: chi2}), but only 
for the total
reconstructed kinetic energy of the final state nucleons. Dotes denote
$\chi^2$ minima.}
\label{fig: kontur}
}
\end{figure}

\begin{figure}
\centering{
\includegraphics[width=\columnwidth]{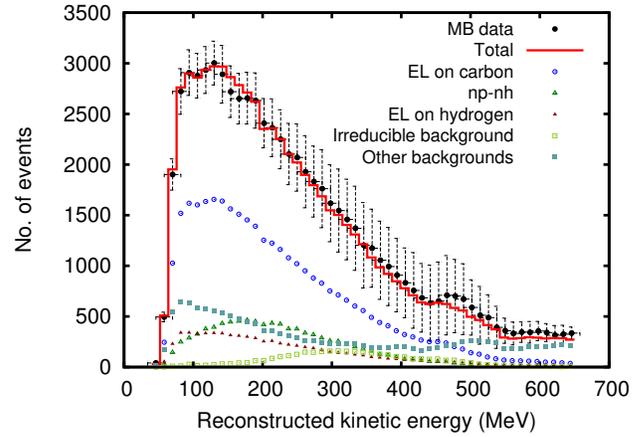}
\caption{[Colour online] The distribution of the total reconstructed kinetic 
energy of the 
final state nucleons, broken down to 
individual contributions from: elastic scattering on carbon, np-nh, 
elastic scattering on hydrogen, irreducible background and other backgrounds. 
The NuWro result is obtained with the $M_A = 1.10$~GeV and $g_A^s = 
-0.4$ values.}
\label{fig: ma_bf}}
\end{figure}

\begin{figure}
\includegraphics[width=\columnwidth]{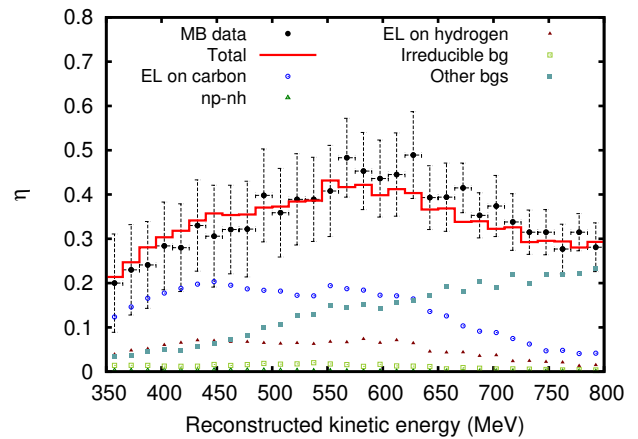}
\caption{[Colour online] The ratio $\eta$ (Eq. \ref{eq: ratio}) as a 
function 
of the total reconstructed kinetic energy of all nucleons in the final 
state, broken down to individual contributions from: elastic 
scattering on carbon, np-nh, elastic scattering on hydrogen, 
irreducible background and other 
backgrounds. The NuWro result is obtained with 
the $M_A = 1.10$~GeV and $g_A^s = 
-0.4$ values.}
\label{fig: ds_bf}
\end{figure}

\subsection{Analysis without $np-nh$ Contribution}

We first repeat the analysis without the $np-nh$ contribution to check if our 
numerical procedures reproduce the MB results.

We assumed a fixed value of the axial mass for hydrogen $M_A = 1.03$~GeV and we 
minimised the $\chi^2$ function for the effective axial mass for carbon 
($M_A^{eff}$) using the data for the reconstructed energy distribution. 
Following the MB procedure we fixed the value $g_A^s = 0$. The value 
$M_A^{eff} = 1.47 \pm 0.10~\mathrm{GeV}$ was obtained with $\chi^2_{min} / DOF = 
23.6 / 50$ (a confidence level (CL) $99.9\%$). This value is larger than the one
reported 
by the MB Collaboration ($M_A^{eff} = 1.39 \pm 0.11 $~GeV) 
\cite{MiniBooNE_AguilarArevalo:2010cx} but consistent within the 1$\sigma$ error 
bars. A discrepancy is probably caused by the presence of the
pionless $\Delta(1232)$ resonance decays in the NUANCE but not in the  NuWro 
generator. Moreover, we use a lower value of $M_A$ for hydrogen than the MB 
Collaboration. 

To investigate a possible impact of the choice of the electromagnetic form 
factors 
parameterisation on the final results, we repeated 
the computations using the form factors  
corrected by a two-photon exchange \cite{Graczyk:2011kh} and obtained almost 
identical results.

Using the data for the ratio $\eta$ (Eq. \ref{eq: ratio}) we examined the 
strange quark contribution to the NCEL cross section. We assumed a fixed value 
of the axial mass: $1.03$~GeV for hydrogen and effective value 
$1.47$~GeV for carbon. We found the strange quark contribution to be  $g_A^s = 0.24 \pm 0.46$ with $\chi^2_{min} / DOF = 26.7 / 29$ (CL $58.8\%$). This results is consistent with values published by the MB Collaboration \cite{MiniBooNE_AguilarArevalo:2010cx} and the BNL E734 experiment \cite{Ahrens:1986xe}.

\subsection{Analysis with $np-nh$ Contribution}
\label{sec: npnh}

Following the same steps the analysis was repeated including the $np-nh$ 
contribution. Assuming $g_A^s = 0$ the minimum of the $\chi^2$ for the 
distributions of the total
reconstructed kinetic energy of the final state nucleons was found for the 
axial mass value $M_A = 
1.15\pm 0.11$~GeV with $\chi^2_{min} / DOF = 24.4 / 50$ (CL $99.9\%$). The 
extraction of the strangeness from the ratio $\eta$, assuming $M_A = 1.15$~GeV, 
leads
to the value $g_A^s = -0.72\pm 0.55$ with $\chi^2_{min} / DOF = 28.7 / 29$ (CL 
$48.1\%$), which is inconsistent with zero (as assumed in the first fit). As 
it was discussed before the two body current events contribute mostly to the 
denominator of the 
ratio $\eta$ making its value smaller. Also, a lower $M_A$ makes the ratio 
$\eta$ smaller. To compensate for both effects a lower value of $g_A^s$ can be 
expected. 

\subsubsection{Simultaneous Extraction of $M_A$ and $g_A^s$}

The inconsistency described in the previous subsection encouraged us to try to 
make a simultaneous fit of both theoretical model parameters. 
In the case of the first observable we obtained the following results:

\begin{itemize}
 \item without $np-nh$  events: $$M_A =  1.34^{+0.18}_{-0.10}\mbox{ 
GeV~~~and~~~} 
g_A^s = -0.5^{+0.2}_{-0.2}$$
with $\chi^2_{min} / DOF = 22.0 / 50$;
 \item with $np-nh$ events: $$M_A =  1.10^{+0.13}_{-0.15} \mbox{ GeV~~~and~~~} 
g_A^s = 
-0.4^{+0.5}_{-0.3}$$
with $\chi^2_{min} / DOF = 22.7 / 50$.
\end{itemize}

In the case of second observable we discovered that the best fit values are 
very sensitive to many details of the theoretical model:

\begin{itemize}
\item from the Fig. \ref{fig: ds_bf} it is clearly seen that the ratio $\eta$ 
depends strongly on ``other backgrounds``;
\item above $350$~MeV of the kinetic energy a significant contribution comes 
from 
irreducible background (pion production and absorption) known with a precision 
not better than 20-30\%;
\item we constructed Monte Carlo $np-nh$ toy models based on the TE model with 
modified distribution of energy transfer and the obtained best fit values depend 
strongly on such modifications; on the other hand the results from the first 
observable are affected in the much weaker way.
\end{itemize}

Fig. \ref{fig: kontur} shows our results for the simultaneous two-dimensional 
fits without and with the $np-nh$ contribution included in the NuWro 
simulations 
together with $68\%$ confidence regions.
The inclusion of the $np-nh$ events makes the best fit result for $M_A$ 
consistent with the world average. It confirms that the difference between 
recent and older axial mass measurements can be explained by taking into account 
two body current contribution. The value of the strange quark contribution 
is found to be consistent with zero.

We calculated the value of $\chi^2$ for the second MB observable with the 
values 
$M_A =  1.10$ and $g_A^s = -0.4$ and obtained $\chi^2 / DOF = 30.2 / 29$. It 
means that the reported values are consistent also with the proton enriched 
sample 
observable.

Our best fit for the distribution of the total 
reconstructed kinetic energy of the final state nucleons and the 
overall NuWro prediction (broken down to individual contributions 
from elastic scattering on carbon, elastic scattering on hydrogen, two body 
current, irreducible background and other backgrounds) is demonstrated in Fig. 
\ref{fig: ma_bf}.
The contribution coming from the two body current amounts to approximately 
$15\%$ 
of the overall distribution affecting both its shape and the normalisation. As 
we focus on the sum of the kinetic energies of all the nucleons in the final 
state, the result is not 
much sensitive to the assumptions made on the $np-nh$ kinematics described in 
Sec \ref{sec: mec}.


Our predictions for the ratio $\eta$ obtained with the values $M_A =  1.10$ and $g_A^s = -0.4$ together with the contributions to the numerator 
coming from various signal events are presented in Fig. \ref{fig: ds_bf}. 

As mentioned in Sec. \ref{sec: nucmodel} the formation zone for the $np-nh$ 
events is assumed to be zero, 
because there is no clear physical motivation to introduce it.
In order to estimate how important the FZ effect can be for the NCEL analysis 
we 
repeated the computations assuming the FZ for the $np-nh$ events to be $1$~fm. 
It turned out that this assumption does not affect the final results in a 
statistically relevant way.

We investigated the impact of the uncertainty of the value of the 
parameter $p_{cc}$ defining a relative abundance of $n-p$ and $n-n$ pairs
on which two body current scattering occur. The default value of $p_{cc}$ is 
$0.6$ and we repeated the computations with the value 
$p_{cc} = 0.9$. No influence on the final results was found.

\section{Conclusions and outlook}
\label{conclusions}

The impact of the two body current events on the analysis of the MB data for 
the neutrino NCEL scattering on $CH_{2}$ was investigated in detail.
This is the first analysis of this kind yet. We performed a simultaneous fit to 
two theoretical model parameters and obtained the values $M_A =  
1.10^{+0.13}_{-0.15}$~GeV and $g_A^s = -0.4^{+0.5}_{-0.3}$.
Our results provide a new evidence that large axial mass measurements can be 
explained by the two-body current contribution to the cross section neglected in 
the experimental data analysis. 

It would be interesting to repeat this analysis using one of 
the microscopic models of the NC two body current 
contribution. Also, it is desirable to include other nuclear effects like random 
phase approximation (RPA)
correlations. As mentioned above, NuWro contains implementations of IFIC and 
Martini et al. models for $np-nh$ in the CC channel only. Recently 
using the theoretical computations from \cite{Graczyk_rpa1} NuWro
was upgraded with the implementation of RPA, but again only for the CC 
reactions. We 
are planning to include $np-nh$ dynamics and the RPA effects for NC
channels but it requires a significant amount of theoretical and programming 
work. For the RPA an appropriate recalculation 
of the components of the polarisation tensor is needed as in the 
relativistic formalism adopted in Ref. \cite{Graczyk_rpa1} the response 
functions 
are calculated in the analytical form.  

Recently the MiniBooNE Collaboration made public the preliminary results from 
the antineutrino NCEL analysis \cite{JoGra}.
They are supplementary to those discussed in this paper.
When the data become available a combined analysis will have more potential to
investigate the $np-nh$ contribution and put more constraints on the $g_A^s$ 
value.

As discussed in Sect. \ref{sec: npnh} the MB proton enriched distribution of 
events contains an important information about hadrons resulting from $np-nh$ 
events not explored in this study. In order to make use of this information a 
very good control over the FSI effects is required.

Finally, we would like to notice that there is an interesting idea for an 
alternative measurement of the NCEL cross section described in 
Ref. \cite{ankowski_prl}. 
The authors investigate the relation between the rate of the observed $\gamma$ 
rays coming from nuclear deexcitation in water-Cherenkov detectors 
and the NCEL cross section.

\section*{Acknowledgments}

We would like to thank Denis Perevalov, Arie Bodek and Eric Christy for useful information. 

The authors were partially supported by the grant No. UMO-2-11/M/ST2/02578.

Most of numerical calculations were carried out in the Wroc{\l}aw Centre for 
Networking and Supercomputing (http://www.wcss.wroc.pl), grant No. 268.

\end{document}